\documentclass[a4paper,11pt,onecolumn]{article}

\usepackage[utf8]{inputenc}
\usepackage[T1]{fontenc}
\usepackage[english]{babel}
\usepackage{graphicx}
\usepackage{subcaption}
\usepackage{booktabs}
\usepackage{geometry}
\usepackage{amsmath}
\usepackage{amssymb}
\usepackage{tabularx}

\usepackage{dsfont}
\usepackage{tikz}
\usetikzlibrary{positioning, fit, arrows.meta, shapes.geometric, backgrounds, calc}

\usepackage{csquotes}
\usepackage[authoryear, round]{natbib}
\usepackage[colorlinks=true, allcolors=darkblue, urlcolor=darkblue]{hyperref}
\usepackage{float}
\usepackage{caption}
\usepackage{siunitx}
\usepackage{parskip}
\usepackage{cite}
\usepackage{titlesec}
\usepackage{authblk}    
\usepackage{hyperref}   
\usepackage{cite}       
\usepackage{lineno}     
\usepackage{xcolor}
\usepackage{microtype}

\definecolor{darkblue}{rgb}{0.0, 0.0, 0.55}

\title{\textbf{Synthesizing the Counterfactual: A CTGAN-Augmented Causal Evaluation of Palliative Care on Spousal Depression}}

\author{Pietro Grassi}
\affil{Sant'Anna School of Advanced Studies, Pisa, Italy}
\date{January 2026}

\titleformat{\section}
{\normalfont\large\bfseries\color{darkblue}}{\thesection}{1em}{}
\titleformat{\subsection}
{\normalfont\normalsize\bfseries}{\thesubsection}{1em}{}
\titleformat{\paragraph}[runin]
{\normalfont\normalsize\bfseries}{}{0em}{}[.]

\titlespacing*{\section}{0pt}{1.5ex plus 1ex minus .2ex}{1ex plus .2ex}
\titlespacing*{\subsection}{0pt}{1.2ex plus 1ex minus .2ex}{0.8ex plus .2ex}

\title{Synthesizing the Counterfactual: A CTGAN-Augmented Causal Evaluation of Palliative Care on Spousal Depression}
\date{\today}

\begin{document}
	
		\begin{center}
			\vspace*{-0.5cm}
			{\Large \textbf{Synthesizing the Counterfactual: A CTGAN-Augmented Causal Evaluation of Palliative Care on Spousal Depression}}
			
			\vspace{0.8cm}
			
			{\small 
				\textbf{Pietro Grassi}\textsuperscript{1},
				\textbf{Roberto Molinari}\textsuperscript{2},
				\textbf{Chiara Seghieri}\textsuperscript{3},
				\textbf{Daniele Vignoli}\textsuperscript{4}
			}
			
			\vspace{0.3cm}
			
			{\footnotesize 
				\textit{\textsuperscript{1}Honours Student at Sant'Anna School of Advanced Studies, Pisa, Italy} \\
				\textit{\textsuperscript{2}Department of Mathematics and Statistics, Auburn University, United States}\\
				\textit{\textsuperscript{3}Management and Health Laboratory, Institute of Management, Sant'Anna School of Advanced Studies, Pisa, Italy} \\
				\textit{\textsuperscript{4}Department of Statistics, Computer Science, Applications ``G. Parenti'', University of Florence, Italy}\\
			}
			
			\vspace{0.8cm}
			
			\begin{minipage}{0.88\textwidth}
				\small \textbf{Summary}
				
				\vspace{0.2cm}
				
				Spousal bereavement severely deteriorates mental health. While palliative care benefits dying patients, its ``stress-buffering'' effect on survivors' depression remains empirically elusive due to acute small-$N$ constraints in longitudinal dyadic data. This study evaluates the causal impact of palliative care on bereaved spouses while introducing Synthetic Data Generation (SDG) to resolve sample attrition in quasi-experimental designs. Using SHARE panel data, we augment the sparse treated cohort via a Conditional Tabular GAN, anchoring synthetic trajectories to empirical baseline constraints to preserve causal pathways. A Matched Difference-in-Differences estimator applied to the high-fidelity augmented dataset evaluates the treatment effect. \\Results reveal a non-linear psychological response. Palliative care initially exacerbates acute depressive symptoms at the time of loss ($\beta_0 = 0.218$, $p < 0.05$), reflecting the intense emotional confrontation of the intervention. However, a sustained stress-buffering effect emerges in subsequent periods ($\beta_2 = -0.763$, $p < 0.01$), indicating an accelerated long-term recovery compared to standard care. Estimates are highly robust to unobserved confounding (Oster’s $\delta > 1$). Substantively, we advocate for reconceptualizing end-of-life care as a dyadic public health intervention. Methodologically, we establish SDG as a robust analytical tool capable of powering fragile quasi-experiments in longitudinal social surveys.
				
				\vspace{0.5cm}
				\noindent \textbf{Keywords:} Causal Inference, Palliative Care, SHARE, Synthetic Data Generation, Widowhood Effect.
			\end{minipage}
		\end{center}
		\vspace{0.8cm}
		\hrule
		\vspace{0.8cm}
	
\section{Introduction}
\label{sec:introduction}

The end of life represents a profound transition not only for the dying individual but also for their immediate social network, particularly their spouse. The bereavement process frequently triggers a cascade of adverse physical and mental health outcomes for the surviving partner, a phenomenon widely documented as the ``widowhood effect'' \citep{christakis2003widowhood, elwert2008effect}. While the direct clinical benefits of palliative care for terminally ill patients are well-established \citep{temel2010early, who2020palliative}, its extended impact (the potential ``ripple effect'' on the psychological well-being of the surviving spouse) remains underexplored. Evaluating whether end-of-life trajectories shaped by palliative care, as opposed to traditional care, mitigate subsequent depressive symptoms in bereaved partners is crucial for designing holistic healthcare policies.

However, estimating this causal effect using observational survey data presents substantial methodological challenges. Randomized controlled trials (RCTs) for end-of-life care are often ethically fraught and practically difficult to implement \citep{wright2008associations}. Consequently, researchers must rely on longitudinal observational datasets, such as the Survey of Health, Ageing and Retirement in Europe (SHARE) \citep{borsch2013data}. While SHARE provides rich, multi-wave data on older adults, causal inference in this context is frequently hindered by acute sample size limitations. The intersection of specific conditions—being observed longitudinally, experiencing the death of a partner between waves, and having clear documentation of the type of end-of-life care received—yields prohibitively small treatment groups. In such small-$N$ settings, traditional quasi-experimental estimators like Difference-in-Differences (DiD) lack the necessary statistical power and variance stability to produce reliable estimates. This inherent data scarcity often renders conclusive causal claims empirically elusive, forcing researchers to seek novel augmentative approaches.

To overcome the limitations of small sample sizes in tabular longitudinal data, this paper extends the emerging intersection of machine learning and causal inference by integrating Synthetic Data Generation (SDG) into a quasi-experimental panel design. While Generative Adversarial Networks (GANs) and Variational Autoencoders (VAEs) have been increasingly adopted in the machine learning literature primarily for privacy preservation \citep{jordon2022synthetic, reiter2005releasing}, their application for data augmentation in quasi-experimental social science research has not been duly explored. For this reason, we employ a Conditional Tabular GAN (CTGAN) \citep{xu2019modeling} to synthetically augment the underrepresented treated and control cohorts. By learning the complex, high-dimensional joint distribution of the real data, the generative model allows us to expand the sample size while faithfully preserving the longitudinal structure, covariate distributions and pre-existing causal pathways of the original SHARE dataset. We subsequently apply a Matched DiD estimator \citep{heckman1997matching, callaway2021difference} to the augmented dataset to estimate the Average Treatment Effect on the Treated (ATT) of palliative care on the surviving partner's depression, measured via the EURO-D scale. By incorporating baseline causal inferential tools (Standard Two-Way Fixed Effects and Dynamic OLS) as benchmarks, we demonstrate the robustness of our preferred matched specification. 

The contribution of this article is therefore twofold. From a clinical perspective, we provide empirical evidence that palliative care significantly mitigates the trajectory of depressive symptoms in bereaved partners compared to traditional end-of-life care. Methodologically, we demonstrate that SDG techniques (specifically CTGAN) offer a viable solution to the small-$N$ constraints prevalent in longitudinal social science research. By validating the structural fidelity of the synthetic data, we show that generative augmentation can power quasi-experimental designs without introducing artificial structural biases. Crucially, rather than altering the underlying causal pathways, the generative model expands the empirical support of the high-dimensional covariate space. This increased density allows matching algorithms to construct highly comparable counterfactuals that were mechanically unattainable in the sparse observational sample. More broadly, our framework offers a scalable blueprint to address the challenges of sub-population sparsity and sample attrition in large-scale longitudinal population surveys, thus opening new avenues for both causal evaluation and broader quantitative analysis in the social sciences.

The remainder of this article is organized as follows. Section \ref{sec:theory} outlines the theoretical framework linking palliative care to survivor well-being. Section \ref{sec:data} describes the SHARE dataset, the SDG procedure, and our matching and DiD methodologies. Section \ref{sec:results} presents the empirical results. Finally, Sections \ref{sec:discussion} and \ref{sec:conclusion} discuss the substantive and methodological implications of our findings and conclude.

\section{Theoretical Framework}
\label{sec:theory}

To contextualize our dual contribution, we first outline the theoretical pathways through which end-of-life care impacts surviving partners. Subsequently, we detail the methodological barriers to estimating these causal effects in survey data and formulate our rationale for employing GANs as an augmentative solution for causal inference and for the broader field of social and health sciences.

\subsection{Palliative Care and the Widowhood Effect: A Stress-Buffering Mechanism}
\label{subsec:theory_clinical}

The transition into widowhood is universally recognized as one of the most stressful life events, consistently associated with elevated risks of morbidity, mortality and substantial psychological distress \citep{stroebe2007health}. Rooted in the Stress Process Model \citep{pearlin1981stress}, the trauma of losing a spouse operates as a primary stressor that triggers secondary psychological consequences, most notably severe depressive symptoms \citep{carr2001late, pang2024depressive}. 

While the trajectory of bereavement is inherently heterogeneous \citep{galatzer2018trajectories}, the context in which the death occurs plays a pivotal role in shaping the survivor's psychological outcomes \citep{carr2001late}. Traditional end-of-life care often involves aggressive, life-prolonging medical interventions within acute hospital settings, which can exacerbate the physical and emotional burden on both the patient and the witnessing partner \citep{wright2008associations}. Conversely, palliative care operates on a fundamentally different paradigm. It is defined by a holistic approach aimed at optimizing the quality of life, managing pain and providing robust psychosocial support, not just to the dying individual, but to their family unit as well \citep{radbruch2020redefining, sallnow2022report}.

From a sociological standpoint, palliative care acts as a critical ``stress-buffering'' mechanism \citep{cohen1985stress, hui2016integrating}. By shifting the focus from curative treatments to comfort and dignity, palliative care facilitates anticipatory grief, allowing the couple to process the impending loss in a controlled, supportive environment \citep{coelho2020family}. Furthermore, the explicit inclusion of the family in the care plan reduces the survivor's feelings of helplessness and caregiver burden \citep{el2017effects}, potentially mitigating the acute onset of depression (EURO-D) following the patient's death \citep{christakis2003widowhood, ornstein2015association}. We hypothesize that surviving partners of decedents who received palliative care will exhibit a less severe trajectory of depressive symptoms compared to those whose spouses received traditional care.

\subsection{The Challenge of Causal Inference in Small-N Longitudinal Settings}
\label{subsec:theory_causal}

Testing the aforementioned hypothesis requires robust causal inference methodologies, typically deployed on rich longitudinal data such as the SHARE dataset. The standard (econometric) approach for evaluating such observational panel data is the Difference-in-Differences (DiD) estimator \citep{roth2023whats, goodman2021difference}, which controls for time-invariant unobserved confounding by comparing the pre- and post-treatment trajectories of treated and control groups \citep{callaway2021difference}.

However, isolating the precise dyadic sub-population required for this design results in severe sample attrition. In the context of the SHARE dataset, this intersectionality frequently reduces the treated cohort to a fraction of the overall sample. In these acute small-$N$ settings, the standard asymptotic properties of quasi-experimental estimators begin to fail, leading to problematic inferential breakdowns. Standard cluster-robust variance estimators with few treated units drastically underestimate standard errors, leading to critical over-rejection of the null hypothesis and a heightened risk of Type I errors (false positives) \citep{ferman2019inference}. Conversely, if researchers apply finite-sample corrections to mitigate this over-rejection, the estimators inherently suffer from a severe loss of statistical power, dramatically increasing the probability of Type II errors and failing to detect true causal effects \citep{mackinnon2018randomization}. Moreover, small samples are highly sensitive to outliers and make it empirically challenging to achieve strict covariate balance or rigorously verify the parallel trends assumption, the foundational requisite for valid DiD estimates \citep{roth2022pretest}. Consequently, navigating this extreme trade-off often forces researchers to either abandon the analysis entirely or accept highly imprecise inferences \citep{kahnlang2020promise}.

\subsection{Data Augmentation for Causal Inference}
\label{subsec:theory_synthetic}

To resolve the tension between theoretical relevance and empirical data scarcity, we integrate Machine Learning (ML) into our causal framework. While Synthetic Data Generation (SDG) has been popularized in computer science and statistics primarily to circumvent privacy restrictions \citep{reiter2005releasing, jordon2022synthetic}, the integration of generative ML into econometric and population survey frameworks represents a frontier methodology. As highlighted by \citet{athey2019machine}, generative architectures like GANs hold immense potential for econometrics by estimating high-dimensional joint distributions and simulating highly realistic datasets. Building on this premise, we postulate that Generative AI—specifically CTGAN \citep{xu2019modeling}—can be explicitly repurposed from a privacy tool into a valid statistical technique for structural data augmentation and sample size expansion in causal designs.

Unlike common imputation methods, CTGANs do not rely on linear assumptions. Instead, they utilize a generator-discriminator architecture to learn possibly highly non-linear, high-dimensional joint probability distributions of the original dataset. By conditioning the generative process on key structural variables, the CTGAN learns to sample from the conditional joint distributions of the panel data. Specifically, rather than generating entirely unconstrained data, we anchor the generative process to an empirical ``static skeleton'', i.e. a vector of fixed, pre-treatment baseline covariates (e.g., gender, baseline wealth, and initial depression score) sampled directly from the real observations, alongside the exact treatment assignment. By conditioning the temporal generation on these observed baseline anchors, the CTGAN can synthesize new, highly realistic longitudinal observations (namely ``digital twins'') that adhere to the empirical marginal distributions of the underlying population. Crucially for causal inference, a well-tuned SDG model preserves the relational structures and conditional probabilities of the original data. This means that the synthesized data retains the inherent selection mechanisms and temporal dynamics (pre-trends) present in the observed SHARE data. By augmenting the small treated cohorts with these high-fidelity synthetic observations, we increase the degrees of freedom, stabilize the variance of the estimators and enhance the statistical power of the Matched DiD model, all while maintaining the structural integrity required for valid causal claims.

\section{Data and Methodology}
\label{sec:data}

To investigate the causal effect of palliative care on the surviving partner's depression trajectory, we employ a multi-step analytical framework. We first isolate a longitudinal sample from a major European social survey. Recognizing the critical sample size constraints inherent in end-of-life dyadic data, we then deploy a generative machine learning architecture to synthesize a high-fidelity augmented cohort. Finally, we apply a Matched DiD estimator to evaluate the causal estimand.

\subsection{Data Source and Analytical Sample}
\label{subsec:data_source}

Data were analyzed using \texttt{R} statistical software\footnote{To ensure reproducibility, the full analytical script is available at \url{https://github.com/pietrograssi-unifi/Synthetic}.}. The primary data source for this study is SHARE, a cross-national, multidisciplinary longitudinal panel database of micro-data on health, socio-economic status and social and family networks of individuals aged 50 or older \citep{borsch2013data}.

Our analytical sample is constructed by linking decedents who died between survey waves to their surviving spouses (the ``survivor''). Inclusion in the final sample requires the survivor to be observed at least one wave prior to the partner's death ($t < 0$) and at least one wave following the death ($t \ge 0$, as the partner's death occurs between $t=-1$ and $t=0$). The treatment variable $D_i$ is a binary indicator derived from end-of-life care retrospective modules: $D_i = 1$ if the decedent received formal palliative/hospice care, and $D_i = 0$ if they received traditional end-of-life care (e.g., standard hospital ward without palliative intent), according to SHARE variable \texttt{xt757\_}: ``In the last four weeks of their life, did the deceased have any hospice or palliative care?''. 

To ensure strict comparability between the treated and control cohorts and to mitigate confounding by indication, we apply a filtering criterion to the control group. For decedents who did not receive palliative care (\texttt{xt757\_} = No), SHARE includes a follow-up question (\texttt{xt754\_}) regarding the reason for this absence. Initially, our extraction identified 9275 survivors whose partners received standard care. However, we restrict our control group ($D_i = 0$) exclusively to survivors who reported that palliative care ``was needed or wanted but not available'' or ``was needed or wanted but too expensive''. Individuals who indicated that such care ``was not needed or wanted'' were systematically excluded from the analytical sample. This critical restriction reduced the counterfactual cohort to a highly comparable final sample of $N = 529$ units. This ensures that the control cohort exhibits end-of-life clinical trajectories and baseline care needs fundamentally similar to the treated cohort, thereby isolating the treatment assignment around exogenous accessibility constraints rather than structural differences in terminal illness severity.

The primary outcome of interest is the survivor's psychological well-being, measured by the EURO-D depression scale \citep{prince1999euro}. The EURO-D is a validated composite index ranging from 0 (no depressive symptoms) to 12 (maximum depressive symptoms), whose robust psychometric properties and cross-cultural measurement invariance within the SHARE longitudinal panel have been extensively confirmed \citep{castro2007ascertaining}. To account for baseline confounding, we extract a rich set of pre-treatment covariates, including demographic (e.g., age, gender), socioeconomic (e.g., wealth, education), and health indicators (e.g., maximum grip strength, limitations in the activities of daily living, survivor's pre-loss depression). To capture the underlying trajectory of the decedent's terminal illness, we computed a binary indicator for the clinical need for palliative care. Following the epidemiological population-based criteria established by \citet{murtagh2014how}, we defined this clinical need as the presence of at least one of the three primary disease categories requiring EoL care: cancer, chronic organ failure (e.g., heart, respiratory, or kidney failure), or progressive neurodegenerative conditions (e.g., Alzheimer's or Parkinson's disease). Given our strict inclusion criteria, the resulting observational sample size is acutely restricted, prompting the need for synthetic augmentation.

\subsection{Synthetic Data Generation: Integrating the CTGAN Architecture}
\label{subsec:method_synthetic}

To overcome the small-$N$ limitation and empower standard causal estimators, we implement a SDG pipeline. The objective of the SDG is to augment the realizations from the observed empirical distributions, thereby stabilizing the variance of our estimators. To systematically select the most appropriate generative model, we benchmarked the CTGAN \citep{xu2019modeling} against other two state-of-the-art architectures: a Tabulare Variational Autoencoder (TVAE) \citep{xu2019modeling} and a customized Transformer-based Tabular VAE (TTVAE) \citep{wang2025ttvae}. The comparative evaluation was formalized leveraging FEST, a unified Framework for Evaluating Synthetic Tabular data \citep{niu2025fest}, which provides a holistic assessment of synthetic data utility and privacy.

The CTGAN is specifically designed to handle the complex modalities of tabular survey data, such as highly skewed continuous variables and imbalanced categorical distributions. However, applying standard generative models to unbalanced longitudinal panel data presents a fundamental architectural challenge, as these networks inherently expect fixed-length, cross-sectional input vectors. To adapt CTGAN for our longitudinal causal design, we engineered a specialized data processing and generation pipeline.

First, the longitudinal dataset was temporally aligned to center on the decedent's death ($t=0$) and reshaped from a long to a wide format. This transformation flattens each survivor's multi-wave trajectory into a single high-dimensional vector, capturing dynamic variables across sequential time steps (e.g., $Y_{t=-1}, Y_{t=0}, Y_{t=1}$). To handle staggered attrition and varying observation lengths, post-exit periods were filled with a standardized padding value, allowing the generative network to implicitly learn the structural bounds of patient survival and study exit.

Second, to ensure rigorous counterfactual comparability and preserve exact baseline distributions, we engineered a partial synthesis strategy \citep{reiter2005releasing} that we refer to as the ``Skeleton Injection'' mechanism. Instead of estimating the full joint distribution, our generative models were trained to learn the conditional temporal distribution of the trajectories given the static baseline characteristics, formulated as $P(Y_{trajectory} \mid X_{skeleton})$. For every synthetic unit generated ($N=3220$ per treatment arm), an empirical skeleton vector ($X_{skeleton}$) was resampled with replacement from the respective real-world cohort. These real structural anchors were then injected into the generation pipeline. Consequently, the generative network synthesized only the time-varying components (the longitudinal trajectories) explicitly conditioned on these exact, real-world static constraints. Importantly, this identical anchoring mechanism was systematically applied across all evaluated architectures (CTGAN, TVAE, and TTVAE) to ensure a fair comparison. The detailed formalization of this partial synthesis mechanism is provided in Appendix \ref{app:skeleton}.

Finally, the synthetic wide data were reconstructed back into a longitudinal long format. During this phase, deterministic evolutionary rules (e.g., age increasing strictly by two years per survey wave) were applied, bypassing the need for the model to learn deterministic temporal laws. Unlike standard GANs, CTGAN employs mode-specific normalization to represent complex continuous variables without suffering from mode collapse. Furthermore, dynamic post-processing clamping rules were applied to guarantee clinical and logical realism: continuous clinical scores (like the EURO-D depression scale) were strictly clipped to their empirical minimum and maximum boundaries and rounded to integer values to perfectly mimic the discrete nature of the original SHARE questionnaire.

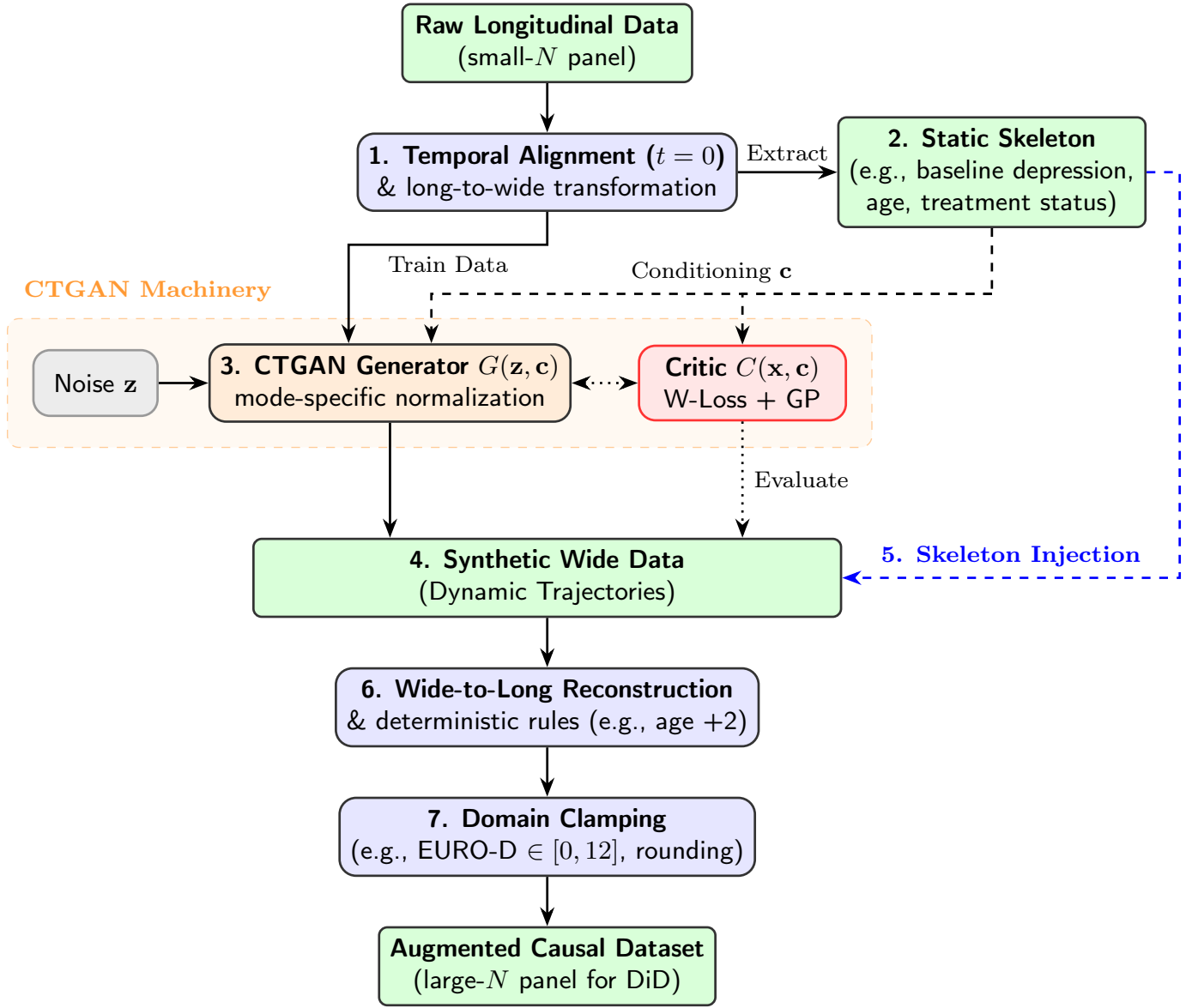
\begin{figure}[htbp]
	\centering
	\resizebox{\columnwidth}{!}{%
		\begin{tikzpicture}[
			>={Stealth[scale=1]},
			font=\sffamily\footnotesize,
			node distance=0.6cm and 0.8cm,
			data/.style={rectangle, draw=black!80, thick, fill=green!15, minimum width=3.5cm, minimum height=0.8cm, align=center, rounded corners=1mm},
			process/.style={rectangle, draw=black!80, thick, fill=blue!10, minimum width=3.5cm, minimum height=0.8cm, align=center, rounded corners=2mm},
			model/.style={rectangle, draw=black!80, thick, fill=orange!15, minimum width=3.5cm, minimum height=0.8cm, align=center, rounded corners=2mm},
			critic/.style={rectangle, draw=red!80, thick, fill=red!10, minimum width=2.5cm, minimum height=0.8cm, align=center, rounded corners=2mm}
			]
			
			\node[process] (wide) {\textbf{1. Temporal Alignment ($t=0$)} \\ \& long-to-wide transformation};
			\node[data, above=0.6cm of wide] (raw) {\textbf{Raw Longitudinal Data} \\ (small-$N$ panel)};
			\node[data, right=1.2cm of wide] (skel) {\textbf{2. Static Skeleton} \\ (e.g., baseline depression,\\ age, treatment status)};
			
			\node[model, below=1.6cm of wide, xshift=-1.9cm] (gen) {\textbf{3. CTGAN Generator} $G(\mathbf{z}, \mathbf{c})$ \\ mode-specific normalization};
			\node[critic, right=0.8cm of gen] (critic) {\textbf{Critic} $C(\mathbf{x}, \mathbf{c})$ \\ W-Loss + GP};
			\node[process, left=0.6cm of gen, fill=gray!15, draw=gray!80, minimum width=1.5cm] (noise) {Noise $\mathbf{z}$};
			
			\node[data, below=1.4cm of gen, xshift=1.9cm, minimum width=7.1cm] (synwide) {\textbf{4. Synthetic Wide Data} \\ (Dynamic Trajectories)};
			\node[process, below=0.6cm of synwide] (long) {\textbf{6. Wide-to-Long Reconstruction} \\ \& deterministic rules (e.g., age +2)};
			\node[process, below=0.6cm of long] (clamp) {\textbf{7. Domain Clamping} \\ (e.g., EURO-D $\in [0,12]$, rounding)};
			\node[data, below=0.6cm of clamp] (final) {\textbf{Augmented Causal Dataset} \\ (large-$N$ panel for DiD)};
			
			\begin{scope}[on background layer]
				\node[fit=(noise)(gen)(critic), fill=orange!5, rounded corners, draw=orange!40, dashed, inner sep=0.3cm] (ctganbox) {};
				\node[above right=0.1cm of ctganbox.north west, font=\bfseries\scriptsize, text=orange!80] {CTGAN Machinery};
			\end{scope}
			
			\draw[->, thick] (raw) -- (wide);
			\draw[->, thick] (wide) -- node[above, font=\scriptsize] {Extract} (skel);
			
			\draw[->, thick] (wide.south) -- ++(0,-0.4) coordinate (train_split) -| node[pos=0.25, below, font=\scriptsize] {Train Data} ([xshift=-0.5cm]gen.north);
			
			\draw[-, thick, dashed] (skel.south) -- ++(0,-0.8) coordinate (cond_split);
			\draw[->, thick, dashed] (cond_split) -| node[pos=0.25, above, font=\scriptsize] {Conditioning $\mathbf{c}$} ([xshift=0.5cm]gen.north);
			\draw[->, thick, dashed] (cond_split) -| (critic.north);
			
			\draw[->, thick] (noise.east) -- (gen.west);
			
			\draw[<->, thick, dotted] (gen.east) -- (critic.west);
			
			\draw[->, thick] (gen.south) -- (synwide.north -| gen.south);
			\draw[->, thick, dotted] (critic.south) -- node[right, font=\scriptsize] {Evaluate} (synwide.north -| critic.south);
			
			\draw[->, thick, dashed, blue] (skel.east) -- ++(0.4,0) |- node[pos=0.75, above, font=\scriptsize, text=blue] {\textbf{5. Skeleton Injection}} (synwide.east);
			
			\draw[->, thick] (synwide) -- (long);
			\draw[->, thick] (long) -- (clamp);
			\draw[->, thick] (clamp) -- (final);
			
		\end{tikzpicture}%
	}
	\caption{End-to-End Synthetic Data Generation Pipeline. The process integrates the core CTGAN machinery within a tailored longitudinal causal framework. After temporal alignment and long-to-wide transformation (1), a static structural skeleton is extracted (2) to condition both the generator and the critic (3). The generated synthetic dynamic trajectories (4) are explicitly anchored back to reality via Skeleton Injection (5). Finally, wide-to-long reconstruction applies deterministic temporal rules (6), and post-processing domain clamping (7) ensures logical and clinical boundaries (e.g., clipping EURO-D scores) before DiD estimation.}
	\label{fig:ctgan_pipeline}
\end{figure}

\subsection{Causal Inference: Matched Difference-in-Differences}
\label{subsec:method_econometrics}

With the augmented dataset, we estimate the ATT of palliative care on the survivor's depression. Even with an expanded sample size, the assignment to palliative care is not random; it is structurally stratified by a complex interplay of clinical and systemic determinants, most notably diagnostic hierarchies (e.g., the well-documented ``cancer privilege''), geographic disparities, and macro-institutional welfare regimes \citep{grassi2026inequalities}. To address this selection bias, we utilize a Matched DiD approach \citep{heckman1997matching, stuart2010matching}.

First, we perform matching on pre-treatment characteristics ($X_{i, t<0}$) to balance the treated and control cohorts. We match individuals based on their baseline EURO-D score, age and socioeconomic indicators, effectively pruning observations lacking common support, i.e. the requisite overlap in the multidimensional distribution of pre-treatment covariates between the treated and control groups. Subsequently, we estimate the causal effect using a generalized Two-Way Fixed Effects (TWFE) dynamic panel specification on the matched sample:\begin{equation}
	\begin{aligned}
		Y_{it} =& \alpha_i + \lambda_t + \sum_{\tau = -K}^{-2} \beta_{\tau} (D_i \times \mathds{1}_{t = \tau}) \\&+ \sum_{\tau = 0}^{M} \beta_{\tau} (D_i \times \mathds{1}_{t = \tau}) + \gamma X_{i,t} + \varepsilon_{i,t},
		\label{eq:did}
	\end{aligned}
\end{equation}where:
\begin{itemize}
	\item $Y_{it}$ is the EURO-D depression score for survivor $i$ at relative time $t$ (centered around the partner's death at $t=0$).
	\item $\alpha_i$ and $\lambda_t$ represent individual and time-period fixed effects, controlling for time-invariant unobserved heterogeneity and common macroeconomic shocks, respectively.
	\item $D_i$ is the treatment indicator (palliative care).
	\item $\mathds{1}_{t = \tau}$ are event-time dummies indicating the wave relative to the death. The wave immediately preceding the death ($t = -1$) is omitted as the reference period.
	\item $X_{i,t}$ represents a vector of time-varying covariates.
	\item $\varepsilon_{i,t}$ is the idiosyncratic error term, clustered at the individual level to account for serial correlation \citep{bertrand2004how}.
\end{itemize}

The coefficients of interest are $\beta_{\tau}$ for $\tau \ge 0$, which capture the post-treatment divergence in depression trajectories between the matched treated and control groups. The coefficients $\beta_{\tau}$ for $\tau < -1$ serve as a formal placebo test to verify the parallel trends assumption: finding these coefficients to be statistically indistinguishable from zero confirms that the synthetic augmentation did not inject spurious pre-treatment divergences.

To validate our findings and demonstrate that the results are not artifacts of a single modeling choice, we benchmark our preferred Matched DiD estimator against alternative econometric specifications, namely a Standard TWFE model (without matching) and a Dynamic OLS estimator. Finally, in the Appendices we complement our primary analysis with a series of robustness checks (specifically, subgroup heterogeneity analyses and an assessment of unobserved confounding via Oster’s bounds).

\section{Results}
\label{sec:results}

The presentation of our empirical findings unfolds in two structural stages. First, we establish the methodological validity of our SDG pipeline. By contrasting pre- and post-generation results, we demonstrate how our constrained CTGAN architecture successfully resolved the substantial lack of common support inherent in the raw SHARE dataset, restoring strict covariate balance. We empirically support the claim that this augmentation operates without injecting structural bias by relying on both algorithmic and econometric foundations. Algorithmically, our conditional generative framework functions as a high-dimensional, non-parametric density estimator: by anchoring the generation to empirical static skeletons, it faithfully reproduces the conditional probability distributions of the observational panel without altering the underlying data-generating process \citep{xu2019modeling, athey2019machine}. Econometrically, the augmented panel successfully passes event-study falsification tests. Following contemporary causal inference standards \citep{roth2022pretest}, the restoration of parallel pre-trends ($\tau \le -2$) serves as empirical evidence that the generative model stabilized the variance and enabled matching without fabricating spurious causal pathways. Second, having validated the causal integrity of the augmented dataset, we deploy our Matched DiD estimator to isolate the causal effect of palliative care on the surviving partner's depression trajectory, revealing a highly dynamic and non-linear psychological response to end-of-life care.

\subsection{Pre-Generation Diagnostics and the Restoration of Common Support}
\label{subsec:Pre-generation}

\begin{figure}[htbp]
	\centering
	\includegraphics[width=0.7\columnwidth]{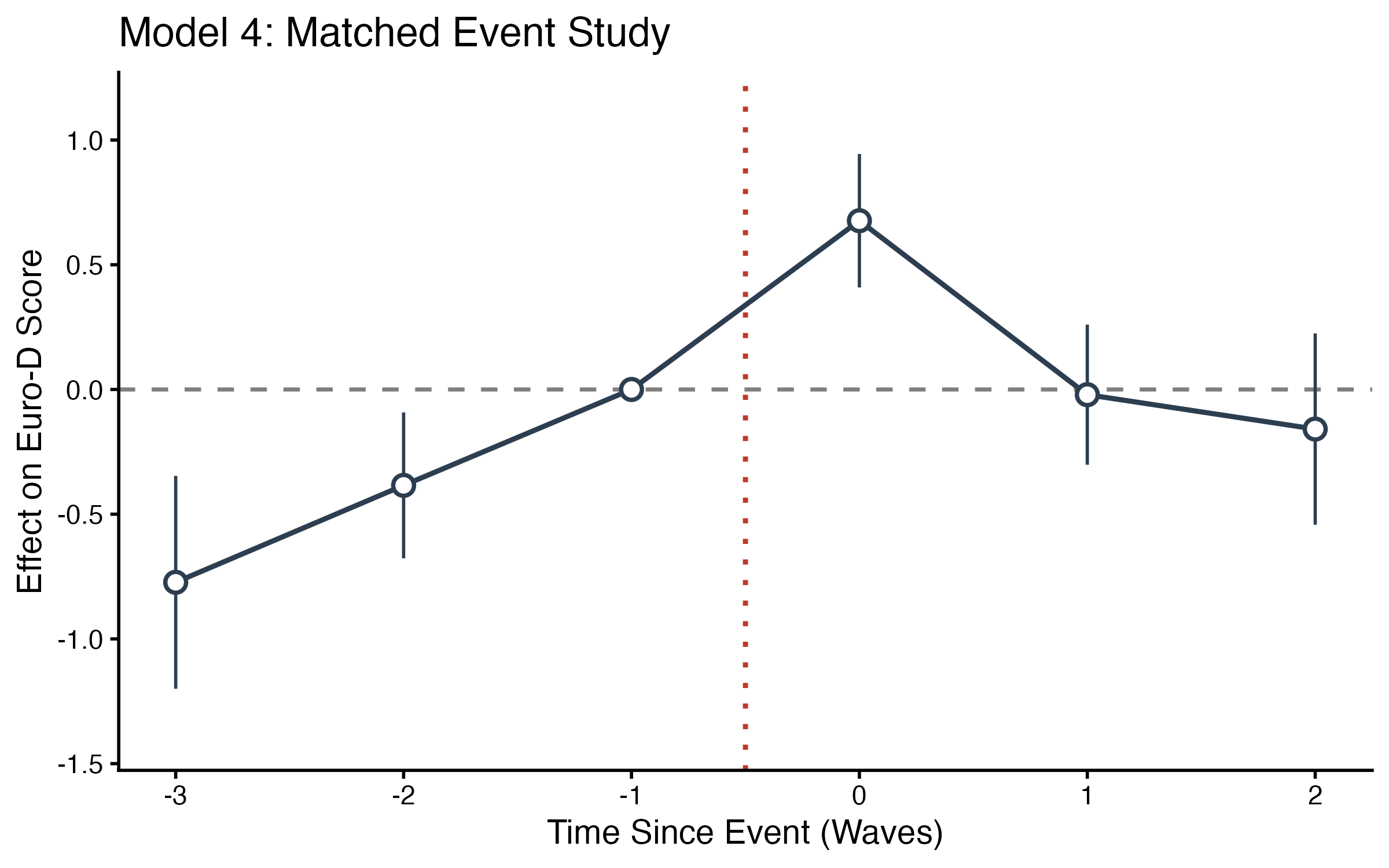}
	\caption{Pre-generation diagnostics: study estimates of the Matched Difference-in-Differences model applied to the raw observational SHARE sample. The statistically significant coefficients in the pre-treatment periods ($\tau \le -2$) highlight the critical violation of the parallel trends assumption, driven by residual selection bias and a fundamental lack of common support in the small-$N$ dataset.}
	\label{fig:M4pre}
\end{figure}

A critical methodological challenge in our observational sample was the severe lack of common support. Due to the small sample size and high covariate imbalance, applying traditional matching estimators directly to the raw SHARE data resulted in poor pair quality, failing to eliminate selection bias and consistently violating the parallel trends assumption (see Figure \ref{fig:M4pre}).

This empirical failure is not an inherent flaw of the DiD design, but a mechanical artifact of matching in small-$N$, high-dimensional settings. Matching estimators suffer from a finite-sample bias driven by the curse of dimensionality: in sparse datasets, the algorithmic distance between the closest available matches inherently grows \citep{abadie2005semiparametric, abadie2006large}. Lacking sufficient common support, the matching algorithm is forced to pair treated units with structurally dissimilar controls in order to minimize the distance function. In a DiD framework, this poor covariate balance is fatal. Because unadjusted baseline characteristics typically drive heterogeneous outcome trajectories over time, this residual confounding mechanically induces divergent pre-trends, ultimately violating the foundational assumption of the estimator \citep{heckman1997matching}. 

By utilizing the CTGAN, we essentially solved this curse of dimensionality. The synthetic augmentation expanded the support of the empirical distribution, populating the covariate space with high-fidelity synthetic twins. As a consequence, the matching algorithm applied to the CTGAN-augmented dataset was able to enforce strict caliper limits—defined as the maximum tolerable mathematical distance allowed between matched units to systematically prevent the pairing of structurally dissimilar observations \citep{rosenbaum1985constructing, austin2011introduction}—and achieve exact covariate balance. This restored the parallel pre-trends, empirically supporting that SDG can successfully rescue causal designs that would otherwise be discarded due to small-sample selection bias.

\subsection{Evaluation of Synthetic Data Generation}
\label{subsec:results_synthetic}

Table \ref{tab:baseline} summarizes the baseline characteristics of the observational SHARE sample before matching and augmentation. Given the strict inclusion criteria for continuous longitudinal observation around the partner's death, the initial sample exhibited severe covariate imbalance and critical small-$N$ limitations (see Table \ref{tab:baseline}), particularly within the treated cohort (palliative care). 

\begin{table}[htbp]
	\centering
	\caption{Baseline Characteristics of the Observational Sample ($t = -1$). Continuous and ordinal variables (e.g., ISCED, SPHUS, ADL) are reported as Mean (SD). Categorical variables are reported as Counts (\%).}
	\label{tab:baseline}
	\resizebox{\columnwidth}{!}{%
		\begin{tabular}{@{}lcccccc@{}}
			\toprule
			& \multicolumn{2}{c}{Palliative Care (N=3220)} & \multicolumn{2}{c}{Standard Care (N=529)} & & \\
			\cmidrule(lr){2-3} \cmidrule(lr){4-5}
			\textbf{Variable} & Mean & Std. Dev. & Mean & Std. Dev. & Diff. in Means & Std. Error \\
			\midrule
			Age (Years) & 75.49 & 9.13 & 73.49 & 8.80 & -2.00 & 0.42 \\
			Education Level (ISCED 0-6) & 2.34 & 1.56 & 2.71 & 1.43 & 0.37 & 0.07 \\
			Self-Perceived Health (1-5) & 3.49 & 1.03 & 3.68 & 1.01 & 0.20 & 0.05 \\
			Baseline EURO-D (0-12) & 3.55 & 2.65 & 3.50 & 2.56 & -0.05 & 0.12 \\
			ADL Score (0-6) & 0.45 & 1.21 & 0.46 & 1.21 & 0.01 & 0.06 \\
			Max Grip Strength (kg) & 26.45 & 9.68 & 26.88 & 10.09 & 0.44 & 0.47 \\
			Log-Wealth & 8.89 & 5.75 & 7.36 & 6.02 & -1.52 & 0.28 \\
			\midrule
			& N & Pct. & N & Pct. & & \\
			\midrule
			\textbf{Gender} & & & & & & \\
			\hspace{1em} Male & 912 & 28.3 & 140 & 26.5 & & \\
			\hspace{1em} Female & 2308 & 71.7 & 389 & 73.5 & & \\
			\addlinespace
			\textbf{Clinical Need for PC} & & & & & & \\
			\hspace{1em} Yes & 1070 & 33.2 & 169 & 31.9 & & \\
			\hspace{1em} No & 2150 & 66.8 & 360 & 68.1 & & \\
			\addlinespace
			\textbf{Cause of Death} & & & & & & \\
			\hspace{1em} Cancer & 1523 & 47.3 & 175 & 33.1 & & \\
			\hspace{1em} Cardiovascular & 759 & 23.6 & 162 & 30.6 & & \\
			\hspace{1em} Respiratory & 193 & 6.0 & 28 & 5.3 & & \\
			\hspace{1em} Other Causes & 745 & 23.1 & 164 & 31.0 & & \\
			\bottomrule
		\end{tabular}%
	}
\end{table}

\begin{figure}[htbp]
	\centering
	\includegraphics[width=0.7\columnwidth]{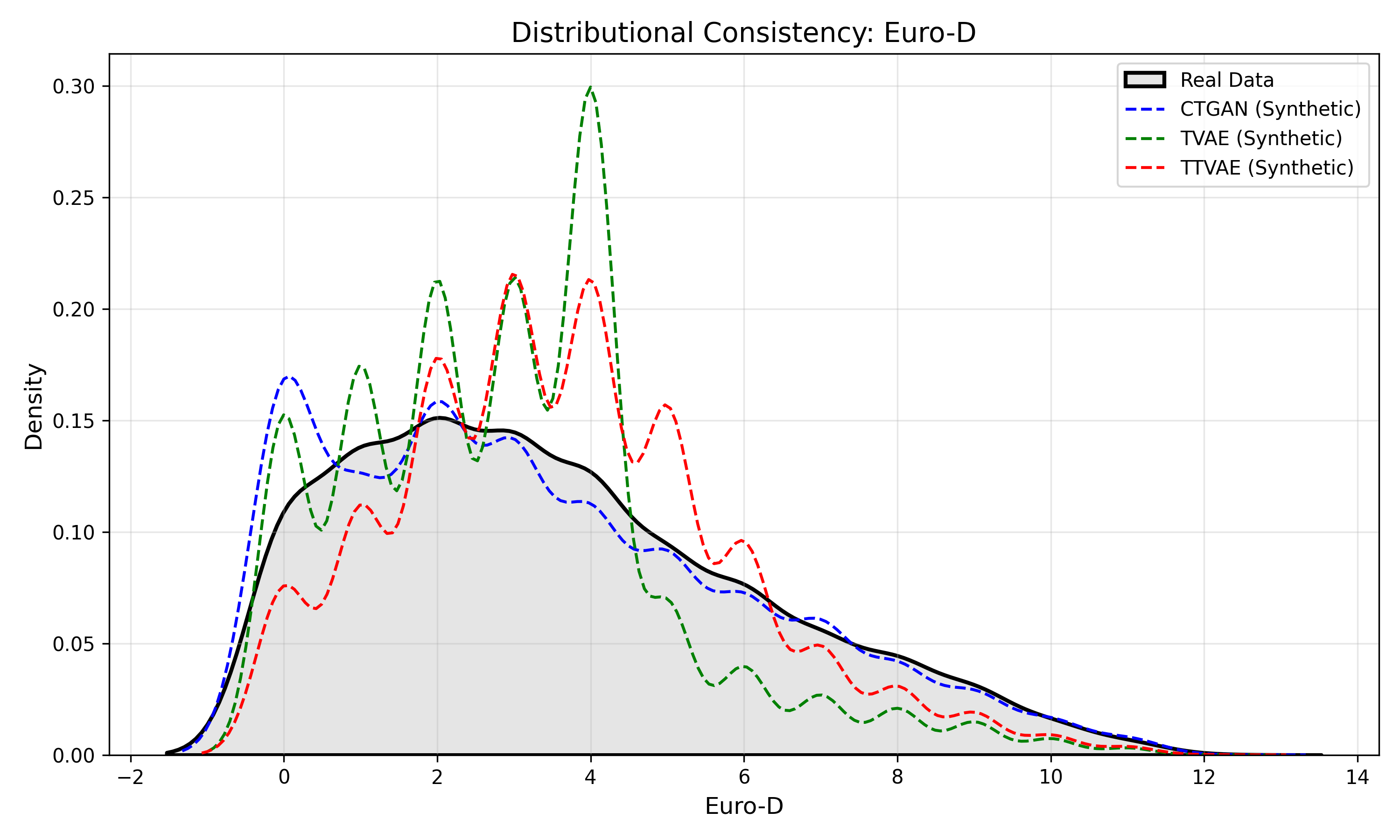}
	\caption{Density distributions of key covariates. The comparison between the original observational data (Real) and synthetic data generated via CTGAN, TVAE, and TTVAE demonstrates CTGAN's superior ability to capture highly skewed clinical variables, such as the EURO-D depression score.}
	\label{fig:dist_comp}
\end{figure}

To validate our synthetic augmentation strategy and systematically select the optimal generative model for causal inference, we benchmarked the CTGAN against two state-of-the-art alternatives (TVAE and TTVAE) using the FEST evaluation framework. While literature frequently associates Variational Autoencoders (VAEs) with oversmoothing artifacts (arising from the mean-seeking nature of their objective functions) our empirical evaluation revealed a different phenomenon. As visually corroborated by Figure \ref{fig:dist_comp}, the VAE-based architectures (TVAE and TTVAE) did not exhibit structural smoothing; rather, they introduced excessive volatility and failed to reliably adhere to the non-normal, right-skewed distribution of the survivor's depression score. Conversely, the CTGAN architecture, leveraging its conditional generator and mode-specific normalization, successfully captured the sharp distributions and the true variance of the observational panel without injecting erratic noise.

This qualitative visual evidence is formally corroborated by the comprehensive suite of FEST evaluation metrics (Table \ref{tab:fest_results}). While the TTVAE architecture exhibited a superior capacity to preserve global pairwise correlations (Correlation Difference = 2.322)—a predictable consequence of the continuous, mean-seeking latent spaces inherent to Variational Autoencoders \citep{kingma2019introduction}—this advantage in linear covariance comes at the cost of marginal accuracy. In our causal design, capturing the non-normal and skewed bounds of discrete clinical variables such as the EURO-D scale is more critical than aggregate correlation matrices. Consequently, our quantitative evaluation confirms that CTGAN emerged as the superior structural model. By prioritizing exact marginal fidelity through mode-specific normalization, the CTGAN achieved an exceptional Kolmogorov-Smirnov (KS) Fidelity Score of 0.937 (outperforming TVAE at 0.905 and TTVAE at 0.860) and minimized the Mean Absolute Percentage Error (MAPE = 0.124).

\begin{table}[htbp]
	\centering
	\caption{Quantitative FEST Evaluation of SDG Architectures}
	\label{tab:fest_results}
	\smallskip
	\footnotesize{\textit{Panel A: Statistical Fidelity and Structural Error}}
	\smallskip
	\begin{tabularx}{\columnwidth}{@{} l *{3}{>{\centering\arraybackslash}X} @{}}
		\toprule
		\textbf{Model} & \textbf{KS Score} ($\uparrow$) & \textbf{MAPE} ($\downarrow$) & \textbf{Corr. Diff.} ($\downarrow$) \\
		\midrule
		\textbf{CTGAN} & \textbf{0.937} & \textbf{0.124} & 3.993 \\
		TTVAE          & 0.860 & 0.137 & \textbf{2.322} \\
		TVAE           & 0.905 & 0.327 & 3.077 \\
		\bottomrule
	\end{tabularx}\\
	\vspace{0.4cm} 
	\smallskip
	\footnotesize{\textit{Panel B: Causal Validity}}
	\smallskip
	\begin{tabularx}{\columnwidth}{@{} l *{2}{>{\centering\arraybackslash}X} @{}}
		\toprule
		\textbf{Model} & \textbf{Adv. Acc.} ($\downarrow$) & \textbf{Pre-Trend Pres.} ($\uparrow$) \\
		\midrule
		\textbf{CTGAN} & 0.643 & \textbf{0.333} \\
		TTVAE          & \textbf{0.573} & 0.238 \\
		TVAE           & 0.626 & 0.272 \\
		\bottomrule
	\end{tabularx}\\
	\vspace{0.4cm}
	\smallskip
	\begin{minipage}{\columnwidth}
		\scriptsize \textit{Note}: KS = Kolmogorov-Smirnov; MAPE = Mean Absolute Percentage Error; Corr. Diff. = Correlation Difference; Adv. Acc. = Adversarial Accuracy; Pre-Trend Pres. = Pre-Trend Slope Preservation. Arrows indicate optimal direction. Optimal values are \textbf{bolded}.
	\end{minipage}\\\vspace{0.4cm}
\end{table}

The viability of a synthetic generator cannot be assessed solely on cross-sectional statistical fidelity; it must be strictly evaluated on its causal validity. A foundational prerequisite for DiD estimation is the parallel trends assumption. Therefore, an augmented dataset that fails to replicate the pre-treatment structural equilibrium inherently invalidates the targeted estimator. 

We formally evaluated the models on their relative ability to preserve the pre-treatment temporal dynamics ($t < 0$). This evaluation underscores a critical methodological distinction: while VAE-based models may excel at smoothing static, cross-sectional correlations, causal identification relies strictly on the preservation of longitudinal trajectories. Although synthesizing noisy, small-sample survey data inherently entails some overall loss of structural fidelity, FEST metrics revealed that CTGAN achieved the highest pre-trend slope preservation score (0.333, compared to 0.272 for TVAE and 0.238 for TTVAE).

Following standard empirical practice in recent DiD literature, we employed event-study falsification tests on the three matched augmented panels as a fundamental plausibility check of the parallel trends assumption \citep{roth2023whats}. Both TVAE and TTVAE failed this pre-treatment falsification, introducing artificial divergences and statistically significant confounding paths before the death event. In contrast, the CTGAN successfully passed this initial plausibility check by minimizing pre-treatment structural distortions. 

However, recognizing that non-significant pre-trends alone does not rigorously guarantee valid causal identification due to inherent risks of low statistical power and pre-test bias \citep{roth2022pretest}, we explicitly complement our CTGAN event-study estimates with formal sensitivity analyses against unobserved confounding, specifically, Oster's bounds, detailed in Appendix \ref{app:oster}. By passing the falsification test where alternative architectures failed, and being fortified by bounding analyses, the CTGAN was uniquely qualified to provide a robust empirical foundation for our design, making it the exclusive choice for our final causal estimations.

\subsection{Causal Estimates: The Effect of Palliative Care}
\label{subsec:results_causal}

\begin{figure}[htbp]
	\centering
	\includegraphics[width=0.7\columnwidth]{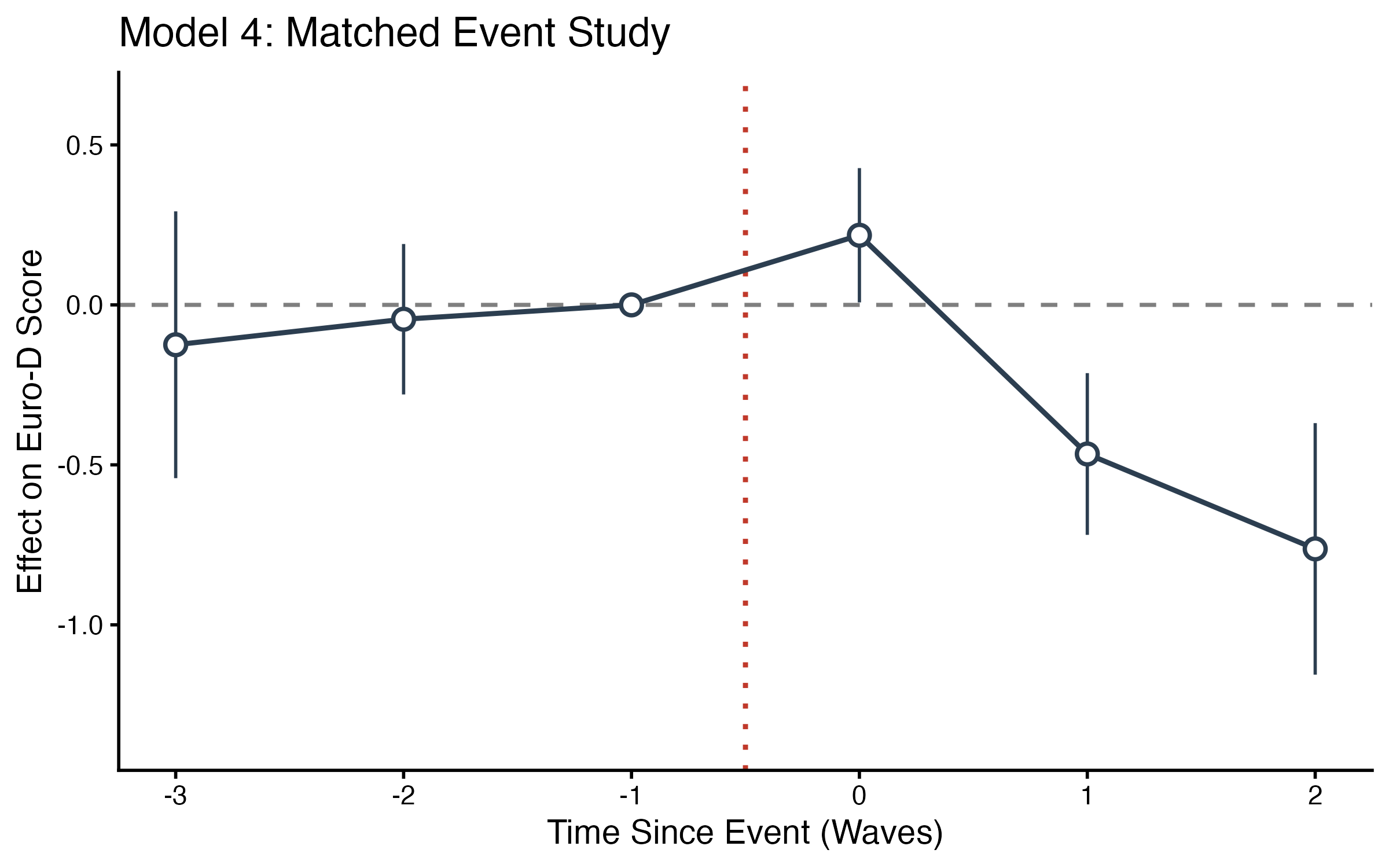}
	\caption{Event-Study Estimates of Palliative Care on Survivor's EURO-D Score (CTGAN + Matched DiD). Error bars represent 95\% confidence intervals.}
	\label{fig:m4_matched}
\end{figure}

Having established the structural integrity of the augmented dataset, we applied our Matched DiD estimator (Equation \ref{eq:did}). Figure \ref{fig:m4_matched} plots the event-study coefficients ($\beta_{\tau}$) representing the ATT at each wave relative to the partner's death ($t=0$). The reference period is the wave immediately preceding the death ($t=-1$).

The event-study plot reveals two critical findings. First, consistent with the successful falsification tests discussed in the previous section, the coefficients for the pre-treatment periods ($t \le -2$) are statistically indistinguishable from zero. This anticipated flat pre-trend confirms that, after matching on baseline covariates, the treatment and control groups were on parallel trajectories regarding their depressive symptoms prior to the end-of-life care intervention. This result not only validates the fundamental assumption of the DiD design, but serves as the definitive empirical corroboration that our CTGAN augmentation successfully preserved the structural dynamics of the original panel without injecting pre-treatment bias. Indeed, the objective function of the generative model is entirely agnostic to the downstream causal estimand; it does not optimize for parallel trends. Had the CTGAN algorithmically hallucinated a protective effect or systematically skewed the treated cohort's trajectories, such structural bias would have mechanically propagated into the pre-intervention periods, thereby violating the falsification test. The presence of parallel pre-trends therefore confirms that the augmentation merely provided the necessary data density for exact matching, faithfully reconstructing the true conditional temporal dependencies of the original observations.

Second, the post-treatment periods reveal a highly dynamic, non-linear causal trajectory. At the wave immediately following the partner's death ($t=0$), we observe a temporary, statistically significant positive spike in the EURO-D score for the treated group ($\beta_0 = 0.218$, $p < 0.05$). Crucially, this positive coefficient indicates that palliative care actually is likely to exacerbate the survivor's acute depressive symptoms in the immediate aftermath of the loss relative to standard care. This initial penalty suggests that the palliative environment (which often encourages open confrontation with mortality, shared decision-making and a deeper emotional involvement in the dying process) may render the immediate emotional shock of bereavement more intense. However, a profound divergence emerges in the subsequent waves. At $t=1$ and $t=2$, survivors whose partners received palliative care exhibit substantially lower depression scores compared to the control group ($\beta_1 = -0.466$, $p < 0.01$; $\beta_2 = -0.763$, $p < 0.01$). 

Because a lower EURO-D score indicates fewer depressive symptoms, these strongly negative point estimates represent a significant positive health outcome for the surviving partner. This non-linear trajectory suggests a complex psychological mechanism: while the intense emotional engagement of palliative care may initially amplify the acute trauma of spousal loss, it ultimately acts as a powerful, sustained ``stress-buffering'' mechanism. By allowing for a more conscious and supported transition during the end-of-life phase, palliative care significantly accelerates and deepens the survivor's long-term psychological recovery.

\begin{table}[htbp]
	\centering
	\caption{Summary of Regression Results across multiple econometric specifications.}
	\label{tab:regression_results_ctgan}
	\begin{tabular*}{\columnwidth}{@{\extracolsep{\fill}}lccc@{}}
		\toprule
		& \textbf{DiD (Base)} & \textbf{Dyn. OLS} & \textbf{Matched DiD} \\
		\midrule
		Treated $\times$ ($t = -3$)         & -0.396$^{\dagger}$ &           & -0.125    \\
		& (0.217)   &           & (0.213)   \\ \addlinespace
		Treated $\times$ ($t = -2$)         & -0.170    & -0.397    & -0.045    \\
		& (0.129)   & (0.271)   & (0.120)   \\ \addlinespace
		Treated $\times$ ($t = 0$) [Event]  & 0.336$^{\S}$ & 0.650$^{\S}$ & 0.218$^{\ddagger}$ \\
		& (0.119)   & (0.192)   & (0.107)   \\ \addlinespace
		Treated $\times$ ($t = +1$)         & -0.228    & 0.632$^{\ddagger}$ & -0.466$^{\S}$ \\
		& (0.175)   & (0.303)   & (0.129)   \\ \addlinespace
		Treated $\times$ ($t = +2$)         & -0.623$^{\ddagger}$ & 0.252     & -0.763$^{\S}$ \\
		& (0.275)   & (0.472)   & (0.200)   \\ \addlinespace
		Lagged Euro-D Score ($t-1$)         &           & -0.453$^{\S}$ &           \\
		&           & (0.020)   &           \\ \midrule
		$R^2$                               & 0.450     & 0.632     & 0.431     \\
		$R^2$ Adj.                          & 0.114     & 0.353     & 0.111     \\
		\bottomrule
		\multicolumn{4}{@{}l}{\footnotesize $^{\dagger} p < 0.1$, $^{\ddagger} p < 0.05$, $^{\S} p < 0.01$} \\
	\end{tabular*}
\end{table}

To ensure these findings are not strictly an artifact of our preferred matching specification, Table \ref{tab:regression_results_ctgan} summarizes the regression coefficients across alternative estimators, including a standard (unmatched) Two-Way Fixed Effects DiD and a Dynamic OLS model. Across the fixed-effects specifications, the core narrative remains remarkably stable: the introduction of palliative care significantly exacerbates the acute depressive spike at the exact time of spousal bereavement ($t=0$), but fosters a deeper psychological recovery in the subsequent periods ($t=1$ and $t=2$). Furthermore, the Dynamic OLS model (which controls for autoregressive depressive trajectories via lagged dependent variables) strongly corroborates the presence of this initial emotional penalty ($\beta_0 = 0.650$, $p < 0.01$). 

The overall consistency of this initial acute shock, coupled with the robust long-term recovery observed in our primary DiD models, underscores the reliability of using high-fidelity synthetic data to support underpowered, high-variance quasi-experiments in longitudinal social science research. Indeed, as explicitly demonstrated in our pre-generation diagnostics (Section \ref{subsec:Pre-generation} and Figure \ref{fig:M4pre}), attempting to estimate this exact causal trajectory using only the raw, small-$N$ observational sample completely failed to achieve common support or satisfy the parallel trends assumption, rendering the analysis empirically intractable without algorithmic augmentation. While the Dynamic OLS specification confirms the initial acute shock, its divergence in later periods is expected. In short-panel settings ($T < N$), models incorporating lagged dependent variables are highly susceptible to Nickell bias \citep{nickell1981biases}, which can distort dynamic treatment effects. Because our Matched DiD estimator explicitly balances pre-treatment covariates and formally satisfies the parallel trends assumption, we retain the Matched DiD as our primary and most unbiased specification for evaluating the long-term causal trajectory.

\section{Discussion}
\label{sec:discussion}

This study was motivated by a dual objective: (i) to evaluate whether palliative care alters the depressive trajectory of surviving partners; and (ii) to demonstrate how SDG can resolve the pervasive small-$N$ constraints inherent in longitudinal survey data. Our empirical results yield significant insights on both fronts.

\subsection{The Non-Linear Trajectory of Grief and Stress-Buffering}
\label{subsec:disc_substantive}

From a clinical perspective, our findings provide robust (see Section \ref{app:oster}), causal evidence that the context of death significantly shapes the trajectory of the widowhood effect. However, contrary to the prevailing expectation that palliative care uniformly dampens depression from the outset \citep{ornstein2015association}, our Matched DiD estimates reveal a dynamic, non-linear psychological response. 

In the immediate aftermath of the loss ($t=0$), surviving spouses of decedents who received palliative care exhibit a statistically significant increase in depressive symptoms ($\approx +0.22$ EURO-D points) compared to those whose partners received traditional care. We interpret this phenomenon as an initial emotional penalty. Indeed, traditional acute care is often characterized by aggressive interventions, opaque prognoses and the marginalization of the family unit, which can paradoxically numb or delay the immediate emotional shock \citep{wright2008associations, sallnow2022report}. In contrast, palliative care intentionally restructures the end-of-life environment to foster open communication, shared decision-making and deep emotional involvement \citep{kuosmanen2021patient, engel2023effective}. By forcing the dyad to explicitly confront mortality, palliative care may render the immediate experience of the death more psychologically intense for the survivor.

However, this initial penalty is the catalyst for a long-term psychological recovery. In the subsequent periods ($t=1$ and $t=2$), the trajectory sharply reverses, with treated survivors scoring up to nearly $0.8$ points lower on the EURO-D scale than their counterfactuals. This significant protective effect strongly aligns with and expands upon the Stress Process Model \citep{pearlin1981stress} and the concept of anticipatory grief \citep{coelho2020family}. By facilitating a good death, palliative care acts as a powerful, delayed stress-buffering mechanism \citep{pokpalagon2025quality}. While existing clinical trials demonstrate that the comprehensive engagement of palliative care equips caregivers with stronger psychological coping mechanisms prior to the loss \citep{el2017effects}, our longitudinal estimates reveal the critical long-term consequence of this preparation: this pre-existing psychological infrastructure ultimately accelerates the survivor's exit from clinical depression in the post-bereavement phase. Consequently, healthcare policies must increasingly view palliative care not merely as an individual medical intervention, but as a dyadic process whose spillover effects generate long-term public health benefits for the bereaved.

\subsection{Synthetic Data Generation as an Epistemological Tool for Population Surveys and Causal Inference}
\label{subsec:disc_methodological}

Beyond the substantive findings, the methodological implications of this paper offer a critical contribution to quantitative social science. The intersectionality required to study end-of-life dyadic transitions in panel data (continuous observation, precise mortality timing, detailed care modalities) inevitably decimates sample sizes. Traditionally, researchers facing such underpowered samples confront a significant methodological trade-off between internal causal validity and statistical power \citep{matthay2020alternative}. In extreme small-$N$ scenarios, this power deficit often forces scholars to either accept highly imprecise inferences or abandon rigorous quasi-experimental designs entirely \citep{kahnlang2020promise}.

By integrating a CTGAN into our causal pipeline, we demonstrate that SDG is a viable epistemological tool to rescue underpowered designs. While ML has predominantly been utilized in econometrics for heterogeneous treatment effect discovery \citep{wager2018estimation, athey2019machine}, our application positions generative models as structural augmenters. Crucially, we argue that the success of generative augmentation in causal inference relies not merely on the underlying neural architecture, but on the rigorous pre- and post-processing framework designed to constrain it. Off-the-shelf application of generative models to panel data frequently fails because these networks (which inherently expect cross-sectional inputs) do not automatically respect longitudinal time-invariance or strict clinical boundaries \citep{yoon2019time}. As detailed in Section \ref{subsec:method_synthetic} (and visualized in Figure \ref{fig:ctgan_pipeline}), our pipeline explicitly enforced causal realism. By introducing a ``Skeleton Injection'' mechanism, we anchored the synthetic dynamic trajectories to real-world baseline clinical and demographic constraints.  Furthermore, applying deterministic post-processing rules (such as domain clamping the EURO-D scores to strict clinical boundaries) provided the necessary epistemological guardrails, forcing the generator to respect the structural laws of the observational survey.

This contextual pipeline explains our most vital methodological warning: the viability of a synthetic generator cannot be assessed solely on cross-sectional statistical fidelity (e.g., KS scores or MAPE), but must be evaluated on its causal validity. While VAE-based architectures distorted the longitudinal dependencies and failed the pre-trend falsification test, the strictly constrained CTGAN successfully preserved the conditional temporal dynamics of the observational data. The flat pre-trends ($\tau \le -2$) observed in our matched event-study model (Figure \ref{fig:m4_matched}) prove that our SDG pipeline did not inject spurious structural breaks. By expanding the support of the empirical distribution within strict logical boundaries, the generative model stabilized the variance of our estimators, yielding precise and statistically significant treatment effects without compromising the underlying causal pathways.

A fundamental methodological paradox arises when applying generative models to underpowered causal designs: if the SDG architecture faithfully replicates the original joint probability distribution, how can it satisfy the parallel trends assumption when the raw observational data failed to do so? This discrepancy is not indicative of an algorithmically induced bias, but rather highlights the distinction between fundamental unobserved confounding and finite-sample mechanical failures. In the sparse original sample, the violation of parallel trends (Figure \ref{fig:M4pre}) was primarily driven by a severe lack of common support; the matching estimator was forced to pair highly dissimilar units, resulting in finite-sample bias. The CTGAN does not alter the underlying data-generating process; rather, it learns it and interpolates the sparse, high-dimensional covariate space. By populating this space with structurally faithful synthetic twins, the generative model provides the density required for the matching algorithm to form exact, high-quality counterfactual pairs. The restoration of parallel pre-trends is therefore the resolution of a small-$N$ matching failure, not an artificial manipulation of the outcome.

Furthermore, we acknowledge that evaluating distributional fidelity metrics (see Table \ref{tab:fest_results}) on acutely small observational samples is inherently noisy. A model might achieve a high statistical fidelity score simply by memorizing a biased small sample. This limitation dictates a critical epistemological shift: in causal inference, the validity of a generative model cannot be proven solely by cross-sectional statistical metrics, but must be corroborated by structural econometric falsification. The fact that our augmented panel demonstrates robustness to unobserved selection (see Section \ref{app:oster}) provides evidence that the CTGAN successfully stabilized the variance without injecting spurious causal pathways.

\subsection{Limitations and Future Directions}
\label{subsec:disc_limitations}

Despite these promising results, several limitations must be acknowledged. Firstly, while our Matched DiD design controls for time-invariant unobserved heterogeneity and balances observable baseline covariates, it cannot entirely eliminate the risk of time-varying unobserved confounding (e.g., sudden, unrecorded shifts in the couple's social support network just prior to the death). Secondly, given the cross-national nature of the SHARE dataset, the observed long-term psychological recovery may be partially moderated by unobserved macro-level heterogeneity. Specifically, structural differences in national welfare policies—such as the availability of subsidized psychological counseling, formal bereavement leave, or dedicated post-loss support systems—vary significantly across European countries and were not explicitly modeled as effect modifiers in our framework. Thirdly, our analysis is constrained by the empirical granularity of the SHARE end-of-life questionnaire. Specifically, the survey aggregates hospice care and at-home care into a single response category. Consequently, our estimates capture the pooled effect of these modalities, preventing us from disentangling the potentially distinct psychological impacts of institutional hospice environments versus home-based palliative support. Fourth, our treatment assignment relies on the retrospective proxy item regarding the deceased's care (variable \texttt{xt754}). While our framework operationalizes this affirmative response as a proxy for a genuine, medically indicated need for palliative care, we cannot definitively confirm the presence of a formal clinical prescription, introducing a potential degree of measurement ambiguity. 

From a methodological perspective, while SDG successfully stabilizes variance, it cannot create new information \textit{ex nihilo}. If an original observational sample suffers from unobserved systemic confounding or fundamental sampling flaws, the generative model will faithfully reproduce and potentially amplify that bias. SDG is an amplifier of empirical distributions, not a corrective mechanism for missing data. In our study, the successful restoration of parallel pre-trends post-generation strongly suggests that the initial imbalance was primarily a mechanical artifact of small-sample variance (lack of common support) rather than unrecoverable selection bias. Nonetheless, this highlights a critical boundary condition for future researchers: the application of SDG requires rigorous pre-processing, specific domain clamping and strict post-generation falsification testing to ensure causal validity.

Future research should explore the theoretical boundaries of generative augmentation in econometric and causal inference frameworks. Specifically, investigating the optimal augmentation factor (i.e., the ratio of synthetic to real data) and its asymptotic properties on the coverage of standard errors remains a critical frontier. Furthermore, applying this CTGAN-DiD framework to other small-$N$ phenomena in the social sciences will help solidify SDG as a standard instrument in the methodologist's toolkit.

\section{Conclusion}
\label{sec:conclusion}

The death of a spouse is a substantial biographical disruption. It precipitates the well-documented ``widowhood effect'', a systemic shock that not only significantly increases the survivor's mortality risk \citep{elwert2008effect} and places them at a significantly heightened risk for severe depressive symptoms, although psychological trajectories vary widely \citep{carr2001late, pang2024depressive}. Evaluating whether the modality of end-of-life care (specifically, palliative versus traditional care) can structurally mitigate this depressive trajectory is a question of paramount importance for public health. Yet, answering this question through observational panel data has historically been impeded by critical sample size constraints, rendering standard quasi-experimental estimators statistically underpowered and highly sensitive to variance. 

In this work we addressed this dual substantive and methodological challenge by integrating SDG into a robust econometric and social survey framework. By employing a CTGAN \citep{xu2019modeling}, we synthetically augmented the underrepresented treated and control cohorts from the SHARE longitudinal dataset. Through FEST evaluation \citep{niu2025fest} and event-study falsification tests, we demonstrated that a properly tuned conditional generative model can learn and reproduce complex, high-dimensional panel data without distorting the underlying causal pathways or violating the critical parallel trends assumption. Applying a Matched DiD estimator to this high-fidelity augmented dataset, our empirical results reveal a non-linear causal effect of palliative care on the bereaved spouse. Results indicate that the emotional confrontation inherent in palliative environments initially exacerbates the survivor's acute depressive symptoms at the time of death. However, this initial psychological penalty subsequently translates into a sustained ``stress-buffering'' effect. In the years following the loss, treated survivors exhibited a marked reduction in depressive symptoms, pointing to an accelerated long-term psychological recovery.

From a public health perspective, our findings substantiate the urgent need to reconceptualize end-of-life care \citep{temel2010early, EAPC25}. Palliative care must no longer be viewed strictly as a patient-centric clinical intervention, but rather as a dyadic, family-centered preventative measure that fundamentally protects the long-term mental health of the surviving partner. In doing so, our empirical framework directly contributes to the growing literature advocating for the systemic integration of palliative care beyond the traditional confines of medicine \citep{knaul2018alleviating, who2020palliative, grassi2026inequalities}. We position high-quality end-of-life care as a foundational pillar of public and social health, a structural intervention essential for mitigating the secondary morbidity of bereavement and safeguarding the social fabric of aging populations. From a methodological standpoint, this work provides a formal blueprint for causal inference in small-$N$ longitudinal settings. We establish that SDG is not merely a mechanism for privacy preservation, but a powerful methodological tool capable of rescuing empirically fragile quasi-experiments. By subjecting generative models to rigorous econometric falsification tests, researchers can bridge the distributional learning power of modern machine learning with the strict identification strategies of causal inference. Ultimately, this integration promises to expand the frontier of quantitative social science, enabling robust policy evaluations for our most vulnerable, hard-to-observe populations.

\section*{Contributors}
All authors conceived and designed the reported study. PG led the data analysis and data augmentation, which was supervised by RM; all authors contributed to the interpretation of results. PG drafted the manuscript; all authors reviewed it critically for important intellectual content. All authors provided final approval of the version to be published. All authors agree to be accountable for all aspects of the work in ensuring that questions related to the accuracy or integrity of any part of the work are appropriately investigated and resolved. PG has directly accessed and verified the underlying data reported in the manuscript. All authors had full access to all of the data in the study and accept responsibility to submit the Article for publication.

\appendix

\section*{Declaration of interests}
The authors declare no conflict of interest.

\section*{Declaration of generative AI and AI-assisted technologies in the manuscript preparation process}
During the preparation of this work the authors used Gemini 3.1 in order to improve readibility of the text. After using this tool/service, the authors reviewed and edited the content as needed and take full responsibility for the content of the published article.

\bibliographystyle{plainnat}
\bibliography{bereaved}

\section*{Code Availability Statement}
The \texttt{R} code used for the data cleaning, multiple imputation, and statistical modeling is openly available at \url{https://github.com/pietrograssi-unifi/Synthesizing-the-Counterfactual}.

\section*{SHARE acknowledgements}
This paper uses data from SHARE Waves 1, 2, 3, 4, 5, 6, 7, 8 and 9  (DOIs:  \url{10.6103/SHARE.w1.900}, \url{10.6103/SHARE.w2.900}, \url{10.6103/SHARE.w3.900}, \url{10.6103/SHARE.w4.900}, \url{10.6103/SHARE.w5.900}, \url{10.6103/SHARE.w6.900}, \url{10.6103/SHARE.w6.DBS.100}, \url{10.6103/SHARE.w7.900}, \url{10.6103/SHARE.w8.900}, \url{10.6103/SHARE.w8ca.900}, \url{10.6103/SHARE.w9.900}, \url{10.6103/SHARE.w9ca900}, \url{10.6103/SHARE.HCAP1.100}) see \citet{borsch2013data} for methodological details.

The SHARE data collection has been funded by the European Commission, DG RTD through FP5 (QLK6-CT-2001-00360), FP6 (SHARE-I3: RII-CT-2006-062193, COMPARE: CIT5-CT-2005-028857, SHARELIFE: CIT4-CT-2006-028812), FP7 (SHARE-PREP: GA N°211909, SHARE-LEAP: GA N°227822, SHARE M4: GA N°261982, DASISH: GA N°283646) and Horizon 2020 (SHARE-DEV3: GA N°676536, SHARE-COHESION: GA N°870628, SERISS: GA N°654221, SSHOC: GA N°823782, SHARE-COVID19: GA N°101015924) and by DG Employment, Social Affairs \& Inclusion through VS 2015/0195, VS 2016/0135, VS 2018/0285, VS 2019/0332, VS 2020/0313, SHARE-EUCOV: GA N°101052589 and EUCOVII: GA N°101102412. Additional funding from the German Federal Ministry of Research, Technology and Space (01UW1301, 01UW1801, 01UW2202), the Max Planck Society for the Advancement of Science, the U.S. National Institute on Aging (U01\_AG09740-13S2, P01\_AG005842, P01\_AG08291, P30\_AG12815, R21\_AG025169, Y1-AG-4553-01, IAG\_BSR06-11, OGHA\_04-064, BSR12-04, R01\_AG052527-02, R01\_AG056329-02, R01\_AG063944, HHSN271201300071C, RAG052527A) and from various national funding sources is gratefully acknowledged (see \url{www.share-eric.eu}).

This paper uses data from the generated easySHARE data set (DOI: \url{10.6103/SHARE.easy.900}), see Gruber et al. (2014) for methodological details. The easySHARE release 9.0.0 is based on SHARE Waves 1, 2, 3, 4, 5, 6, 7, 8 and 9 (DOIs: \url{10.6103/SHARE.w1.900}, \url{10.6103/SHARE.w2.900}, \url{10.6103/SHARE.w3.900}, \url{10.6103/SHARE.w4.900}, \url{10.6103/SHARE.w5.900}, \url{10.6103/SHARE.w6.900}, \url{10.6103/SHARE.w7.900}, \url{10.6103/SHARE.w8.900}, \url{10.6103/SHARE.w9.900}).

\newpage

\setcounter{table}{0}
\setcounter{figure}{0}
\renewcommand{\thetable}{A\arabic{table}}
\renewcommand{\thefigure}{A\arabic{figure}}
	
	\section{The Skeleton Injection Mechanism}
	\label{app:skeleton}
	
	A fundamental challenge when applying SDG to quasi-experimental designs is avoiding the fabrication of baseline confounding. If a generative model slightly hallucinates the marginal distribution of critical pre-treatment variables (e.g., baseline wealth or age), subsequent propensity score matching algorithms will pair synthetic units based on artificial topologies, compromising the causal estimand \citep{amad2025improving}.
	
	To circumvent this, we adopted a partial synthesis framework \citep{reiter2005releasing}, strictly dividing our high-dimensional wide vectors into two distinct feature spaces: (i) a static skeleton $\mathbf{X}$ comprising immutable pre-treatment variables, including demographic anchors (age, gender), fixed socioeconomic status, decedent’s cause of death, and the baseline (pre-loss) EURO-D depression score, and (ii) time-varying vectors $\mathbf{Y}$ representing the sequence of outcomes and changing states across the observed periods ($t-3, t-2, t = 0, t+1, t+2$).
	
	Standard unconstrained GANs attempt to learn the full joint density $P(\mathbf{X}, \mathbf{Y})$. While computationally elegant, this process inevitably introduces sampling variance into $\mathbf{X}$. Our Skeleton Injection mechanism mathematically restricts the generator's task to learning the conditional distribution $P(\mathbf{Y} \mid \mathbf{X})$. The generative algorithm proceeds as follows:
	\begin{enumerate}
		\item Let $N_{target}$ be the desired augmented sample size for a given treatment arm. We draw a sample of size $N_{target}$ with replacement from the raw, empirical distribution of skeletons, yielding a structurally perfect baseline matrix $\mathbf{X}_{emp}$.
		\item The generative model receives $\mathbf{X}_{emp}$ as a conditional input matrix. For each empirical row vector $x_i \in \mathbf{X}_{emp}$, the generator synthesizes the corresponding dynamic trajectory $\tilde{y}_i$, conditionally bound to the constraints of $x_i$.
		\item The final synthetic wide matrix is constructed by concatenating the exact empirical skeletons with their conditionally generated synthetic trajectories $[\mathbf{X}_{emp}, \mathbf{\tilde{Y}}]$.
	\end{enumerate}
	
	This anchoring process ensures that the fundamental cross-sectional density required for propensity score matching at $t=-1$ remains strictly faithful to the original data-generating process, leaving the deep learning architecture exclusively responsible for interpolating the missing longitudinal transitions.
	
\section{Heterogeneity Analyses and Sample Size Limits}
	\label{app:heterogeneity}
	
	The Average Treatment Effect on the Treated (ATT) reported in our main findings provides a pooled estimate of the intervention. To explore potential demographic, clinical, and institutional boundaries, we re-estimated our Matched DiD model across multiple subgroups, explicitly stratifying by gender and European welfare regimes. For each subgroup, we strictly re-calculated the propensity score matching on the baseline wave ($t=-1$) to ensure common support within the restricted strata. 
	
	\begin{figure}[htbp]
		\centering
		\includegraphics[width=\columnwidth]{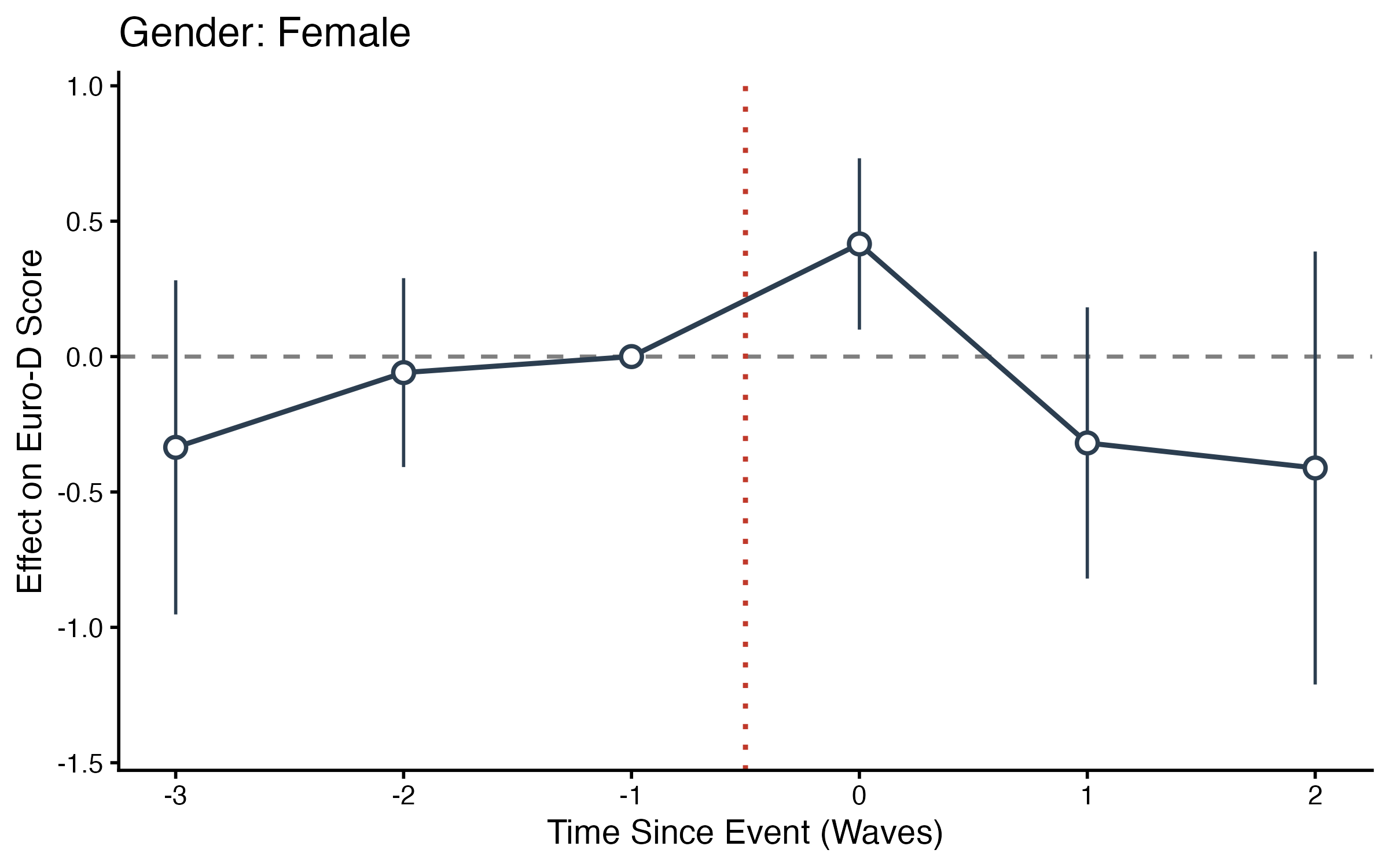}
		\includegraphics[width=\columnwidth]{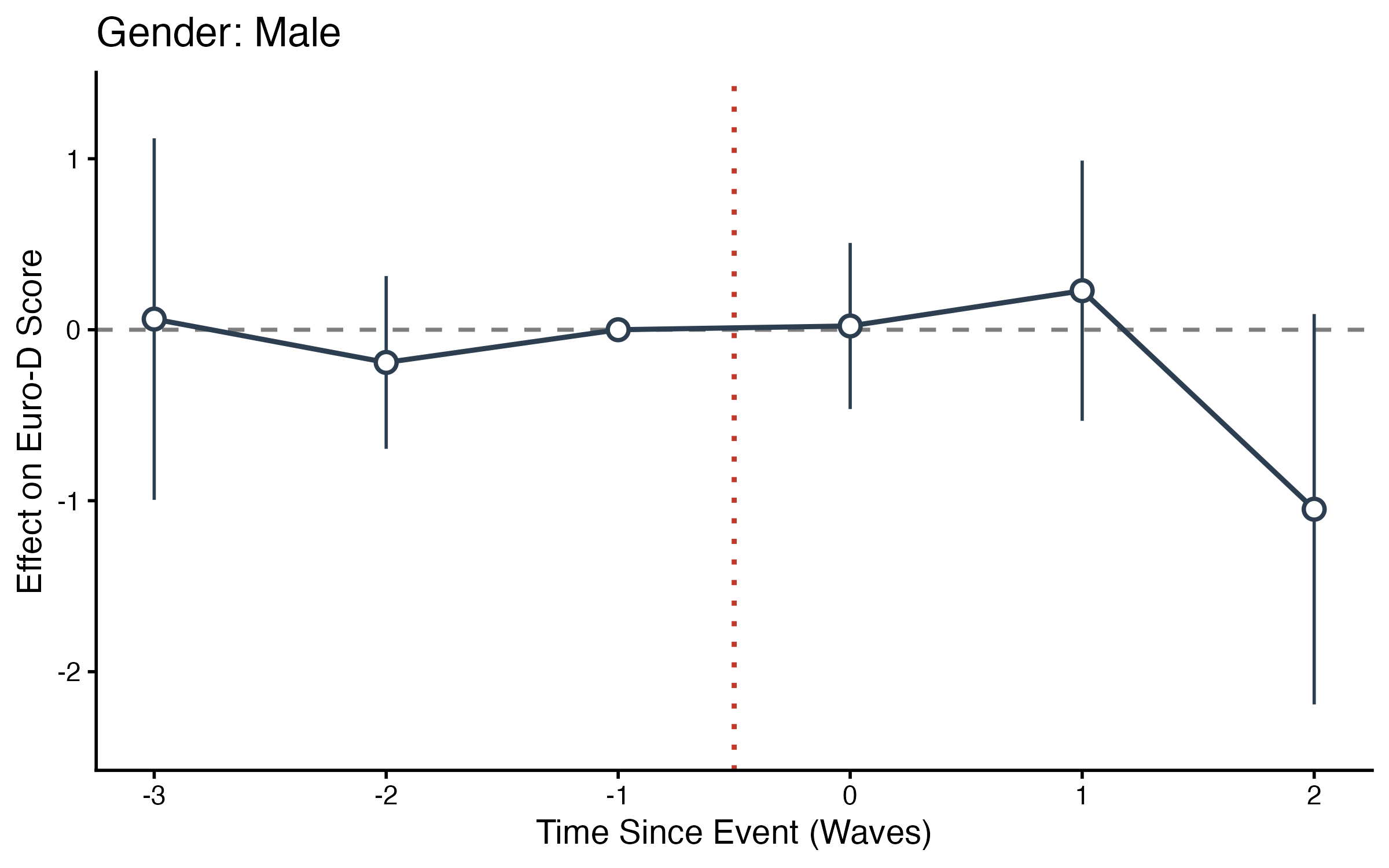}
		\caption{Heterogeneity by Gender. Event-Study Estimates of Palliative Care on Survivor's EURO-D Score for Female (above) and Male (below) sub-samples.}
		\label{fig:app_gender}
	\end{figure}
	
	\begin{table}[htbp]
		\centering
		\caption{\textbf{Heterogeneity by Welfare Regime}}
		\label{tab:app_welfare}
		\resizebox{\columnwidth}{!}{%
			\begin{tabular}{lcccccc}
				\toprule
				Subgroup & t = -3 & t = -2 & t = -1 (Ref) & t = 0 & t = +1 & t = +2 \\
				\midrule
				\textbf{Continental} &  0.162 (0.679) &  0.115 (0.354) & Ref. & 0.167 (0.349) & -0.611 (0.565) & -1.153 (0.88)  \\ \addlinespace
				\textbf{Eastern}     & -0.079 (0.57)  & -0.013 (0.273) & Ref. & 0.389$^{\dagger}$ (0.234) & -0.185 (0.36)  & -0.575 (0.61)  \\ \addlinespace
				\textbf{Southern}    & -0.687 (0.588) & -0.456 (0.298) & Ref. & 0.676$^{\ddagger}$ (0.272) &  0.357 (0.453) & -0.071 (0.694) \\ \addlinespace
				\textbf{Nordic}      & -0.522 (0.923) & -0.287 (0.49)  & Ref. & 0.061 (0.481) & -0.714 (0.867) & -0.485 (1.382) \\
				\bottomrule
				\multicolumn{7}{l}{\footnotesize $^{\dagger} p < 0.1$, $^{\ddagger} p < 0.05$} \\
			\end{tabular}%
		}
	\end{table}
	
	As illustrated in the visual example of gender (Figure \ref{fig:app_gender}) and formally detailed for institutional strata in Table \ref{tab:app_welfare}, the point estimates across most subgroups remain directionally consistent with the non-linear trajectory of our primary pooled findings—displaying an initial depressive penalty at $t=0$ followed by a protective long-term recovery at $t>0$. 
	
	However, the extreme stratification of the augmented sample inherently decimate the degrees of freedom. By slicing the data into micro-strata the standard errors inevitably inflate, frequently rendering the strata-specific post-treatment estimates statistically indistinguishable from zero.
	
	Rather than invalidating the main pooled effect, this loss of statistical power in the partitioned models forcefully underscores our central methodological premise: the structural complexity of matched panel designs requires massive sample support. Even with SDG, disaggregating the augmented cohort rapidly reintroduces the small-$N$ penalty. This dynamic serves as a vital empirical demonstration of the limits of causal inference when applying rigorous quasi-experimental estimators to tightly subgrouped rare longitudinal events.
	
	\section{Assessment of Selection on Unobservables (Oster's Bounds)}
	\label{app:oster}
	
	While our matching procedure effectively balances observable covariates and restores parallel pre-trends, it cannot mathematically account for selection on unobservables (e.g., unrecorded family dynamics or sudden localized economic shocks). To quantify the robustness of our estimates to potential hidden bias, we calculate bounds for the treatment effect following the methodology proposed by \citet{oster2019unobservable}. 
	
	This approach assesses coefficient stability relative to movements in $R^2$ when observable controls are included. Assuming that selection on unobservables is proportional to selection on observables, we calculate Oster's $\delta$ and the identified bound. We apply the standard assumption in the econometric literature that the maximum theoretical $R^2$ is $R_{\max}=1.3\tilde{R}^2$.
	
	\begin{table}[htbp]
		\centering
		\caption{\textbf{Robustness: Oster's Bounds Analysis (Matched DiD)} \\ Assessing Sensitivity to Unobserved Selection}
		\label{tab:app_oster_bounds}
		\begin{tabular*}{\columnwidth}{@{\extracolsep{\fill}}lcccc@{}}
			\toprule
			Period & Estimate & $R^2$ Movement & Bounds & Stability ($\delta$) \\
			\midrule
			\textbf{t = 0 (Death)} &  0.307 & 0.45 $\rightarrow$ 0.45 & [0.307, 0.307]   & $> 10$ (Very Robust) \\ \addlinespace
			\textbf{t = +1}        & -0.225 & 0.45 $\rightarrow$ 0.45 & [-0.225, -0.225] & $> 10$ (Very Robust) \\ \addlinespace
			\textbf{t = +2}        & -0.637 & 0.45 $\rightarrow$ 0.45 & [-0.637, -0.637] & $> 10$ (Very Robust) \\
			\bottomrule
		\end{tabular*}
	\end{table}
	
	As shown in Table \ref{tab:app_oster_bounds}, the computed bounds for the post-treatment coefficients strictly exclude zero, and the calculated $\delta$ values substantially exceed the conventional safety threshold of $1.0$, suggesting that unobserved covariates would need to be more than exactly as important as the included covariates to explain away the effect. This indicates that unobserved confounders would need to be highly influential—substantially more so than our rich set of observed socio-demographic and clinical covariates—to nullify the estimated stress-buffering effect of palliative care.
	
\end{document}